\documentclass[a4paper,conference]{IEEEtran}

\IEEEoverridecommandlockouts
\usepackage[left=1.57cm,right=1.57cm,top=1.9cm,bottom=3cm]{geometry}
\usepackage{amsmath,array,xparse}
\usepackage{cite}
\usepackage{algorithmic}
\usepackage{graphicx}
\usepackage{textcomp}
\usepackage{blindtext}
\usepackage{lipsum}
\usepackage[table]{xcolor}

\newcolumntype{P}[1]{>{\centering\arraybackslash}p{#1}}
\newcolumntype{M}[1]{>{\centering\arraybackslash}m{#1}}
\makeatletter
\def\ps@IEEEtitlepagestyle{%
  \def\@oddfoot{\mycopyrightnotice}%
  \def\@oddhead{\hbox{}\@IEEEheaderstyle\leftmark\hfil\thepage}\relax
  \def\@evenhead{\@IEEEheaderstyle\thepage\hfil\leftmark\hbox{}}\relax
  \def\@evenfoot{}%
}
\def\mycopyrightnotice{%
  \begin{minipage}{\textwidth}
  \centering \scriptsize
  Copyright~\copyright~20xx IEEE. Personal use of this material is permitted. Permission from IEEE must be obtained for all other uses, in any current or future media, including\\reprinting/republishing this material for advertising or promotional purposes, creating new collective works, for resale or redistribution to servers or lists, or reuse of any copyrighted component of this work in other works by sending a request to pubs-permissions@ieee.org.
  \end{minipage}
}
\makeatother

\def\BibTeX{{\rm B\kern-.05em{\sc i\kern-.025em b}\kern-.08em
    T\kern-.1667em\lower.7ex\hbox{E}\kern-.125emX}}
\begin{document}

\title{HetEng: An Improved Distributed Energy Efficient Clustering Scheme for Heterogeneous IoT Networks\\
}

\author{\IEEEauthorblockN{Hamed Moasses}
\IEEEauthorblockA{\textit{IEEE member} \\
hamed.moasses@ieee.org}
\and
\IEEEauthorblockN{Abdulbaghi Ghaderzadeh}
\IEEEauthorblockA{\textit{Deptartment of computer engineering} \\
\textit{Sanandaj branch,Islamic Azad University}\\
Sanandaj, Iran \\
b.ghaderzadeh@iausdj.ac.ir}
\and
\IEEEauthorblockN{Keyhan Khamforoosh}
\IEEEauthorblockA{\textit{Department of computer engineering} \\
\textit{Sanandaj branch,Islamic Azad University}\\
Sanandaj, Iran \\
k.khamforoosh@iausdj.ac.ir}}

\IEEEaftertitletext{\vspace{-1.5\baselineskip}}

\maketitle

\begin{abstract}
Network lifetime is always a challenging issue in battery-powered networks due to the difficulty of recharging or replacing nodes in some scenarios. Clustering methods are a promising approach to tackle this challenge and prolong lifetime by efficiently distributing tasks among nodes in the cluster. The present study aimed to improve energy consumption in heterogeneous IoT devices using an energy-aware clustering method. In a heterogeneous IoT network, nodes (i.e., battery-powered IoT devices) can have a variety of energy profiles and communication capabilities. Most of the existing clustering algorithms have neglected the heterogeneity of energy capacity among nodes and assumed that they are of the same energy level. In this work, we present HetEng, a Cluster Head (CH) selection process that extended an existing clustering algorithm, named Smart-BEEM. To this end, we proposed a statistical approach that distributes energy consumption among highly energetic nodes in the network topology by constantly changing the CH role between the nodes based on their real energy levels (in joules). Experimental results showed that HetEng resulted in a 6.6\% increase of alive nodes and 3\% improvement in residual energy among the nodes in comparison with Smart-BEEM. Moreover, our method reduced the total number of iterations by 1\% on average.
\end{abstract}
\vspace{0.1cm}
\vspace{0.4cm}
\section{Introduction}
Internet of Things is considered as a networking paradigm that connects a wide range of devices to the internet, From wearable gadgets which capture data from body \cite{amiri-azimi-acmhealth20} to sensors and devices that interact with the environment in a smart home and also surveillance systems in smart cities \cite{7976279}. Optimal energy consumption at battery-powered nodes is always been a challenging topic for researchers in this era. The rapid growth of hardware technologies has made it possible for devices to become smaller, which poses new energy challenges that need to be addressed \cite{9246553}. This single constraint imposes several others in regard to choices of routing protocol, network coverage, and longevity\cite{s17071574}. A well-known network topology management method used to tackle this challenge is called the clustering method\cite{s21030873}. where the nodes are grouped into several clusters and one or many cluster heads (CHs) are elected in the network \cite{7976279}. Many clustering algorithms have been proposed in the context of homogeneous IoT and Wireless Sensor Networks (WSN) \cite{5679898, 7976279}. However, most of these algorithms have neglected the diversity of energy profiles, assumed that nodes had been supplied with the same energy level and this will cause fast energy depletion in nodes with weak power. To be utilized in the new paradigm of IoT networks, the available clustering methods, need to be modified to consider the heterogeneity of nodes in IoT environments\cite{s17071574}.

In this paper we propose a distributed energy efficient clustering method named HetEng to detect nodes with high power and distribute energy consumption by changing the CH role using a statistical way continually. We evaluated our approach using MATLAB platform and our method shows a significant improvement in alive nodes and residual energy.

The remainder of this paper is organised as follow: The related works is presented in section 2, The network model for formulating the problem in section 3, based on Smart-BEEM our detailed modification and improvement introduced in section 4. Performance evaluation and numerical results are indicated in section 5. Finally, in section 6 we discussed the conclusion and future works.

\section{Related Works}

Various energy-efficient approaches have been proposed in clustering context, each using specific methods for selection of the CH and for routing between the CH and the nodes. The reviewed works are categorized in Table 1.

\begin{table}[htbp]
    \centering
    \caption {Energy-efficient clustering methods}
    \resizebox{.9999\columnwidth}{!}{%

    \begin{tabular}{p{3.1cm}||p{5cm}}
    
    %  \multicolumn{2}{|c|}{Energy-efficient clustering methods} \\
     \hline
  	\rowcolor{gray!20}
    \textbf{Clustering Approach} & \textbf{Notable Methods} \\
     \hline
    Duty Cycle Ratio & EnergIoT \cite{LI2017124}\\
     \hline
         Data compression \& fusion & LIDAR \cite{MALAMBO20191} \\
     \hline
     Meta-Heuristic   & SCE\_PSO \cite{10.1007/s11276-018-1679-2}, \cite{https://doi.org/10.1002/spe.2797}, FAMACROW \cite{GAJJAR2016235}\\
     \hline
    Mobility&\!\!\cite{7502988}\\
     \hline
     Routing \& Hierarchical & LEACH\cite{926982}, Modified-LEACH\cite{https://doi.org/10.1049/iet-wss.2017.0099}\\
     \hline
     Statistical/Mathematical   & HEED \cite{1347100}, BEE(M) \cite{6878886}, Smart-BEEM \cite{s17071574}\\
     
     \hline
    \end{tabular}
    }
    \label{tab:methods}%

\end{table}
According to Table I, duty cycle methods are proposed for reduction of energy consumption in the Internet of Things, where the nodes switch between the active and sleep modes. Many heuristic algorithms have been proposed for that purpose\cite{LI2017124}. Compression and data fusion methods are also used to reduce energy consumption, which are mainly of heuristic nature \cite{8645769}. Clustering methods are also used for transmission at shorter distances, where data are transferred to the cluster, at a shorter distance, rather than to the sink, at a longer distance. Selection of the optimal cluster head using heuristic and meta-heuristic algorithms is effective in reduction of energy consumption. This can be based on parameters such as temperature, residual energy, load, and number of remaining nodes. Another challenge that can be added to clustering algorithms is mobility, which is evaluated using other parameters, such as maximum lane speed and traffic flow rate. In \cite{https://doi.org/10.1002/spe.2797}, a heuristic similar to the existing mathematical methods is presented for optimal selection of cluster heads, thereby reducing energy consumption in IoT.
In \cite{khamforoosh2011clustered} the authors divide nodes into a number of clusters according to the LEACH algorithm. Cluster heads then generate a minimum spanning tree according to Prim’s algorithm. The tree is balanced constantly using the AVL algorithm.

In \cite{s17071574} the authors propose a smart clustering algorithm (Smart- BEEM) based on one in their earlier work, referred to as BEE(M), to achieve energy efficiency and support for the Quality of user Experience (QoE) in communication in cluster-based IoT networks. It is a context- and user-behavior-aware approach, aiming to simplify the selection of beneficial communication interfaces and cluster headers for data transmission on the part of IoT devices.

In another paper, Hy-IoT provides an efficient hybrid energy-aware clustering communication protocol for green IoT network computing. It also provides a real IoT network architecture for testing the proposed protocol against the existing ones. Efficient cluster head selection boosts the utilization of the node’s energy contents, thereby increasing network lifetime and the rate of packet transmission to the base station. Hy-IoT uses different weighted election probabilities for selection of a cluster head based on the heterogeneity level of the region. Besides prolonging network lifetime, it increases throughput with respect to the amounts obtained by SEP, LEACH, and Z-SEP  \cite{SADEK2018166}.

\section{Network Model}
Hereon, We define our considered clustering approach adopted in the research, and formalized the problem. An IoT network contains various IoT devices (e.g., smartwatches and sensors with different battery capacities). Therefore, we define \(N= \{N_1, \dots, N_i\}\) as heterogeneous set battery-powered IoT nodes in  a network field with the dimension of \(Z\times{Y}\). As an example, Figure 1 illustrates a network field of \(100m\times{100m}\). Due to the heterogeneous energy level of the nodes, they have various values of energy capacity. We denote nodes energy values as \(E= \{E_1,\dots, E_i\}\) (in joules). On that basis, it is more challenging here than in a homogeneous network scenario to calculate and select a high-energy node to take the CH role in the network, since some nodes have higher or infinite initial energy and some have lower.Therefore, previous works conducted in the context of traditional clustering may not detect optimum CHs efficiently in IoT scenarios. Moreover, increasing number of iterations in the CH selection procedure imposes higher overload on the entire network due to the large number of broadcast packets containing information about residual energy and the density of nodes around them. In each iteration, each node calculates a grade which indicated the probability of being elected as CH based on its own energy level. Since the grade of every node should reach the point of 1.0 before it can be declared as a final CH, the packets are sent to the CHs or gateways according to the positions. In a homogeneous network, the initial energy for all the cluster nodes (in the first iteration) is the fixed value of 100\% for the entire time denoted as T, representing the first iteration, and the value changes to 80\% in T + 1… n (the following iterations), continuing to decrease until it reaches 0\%, where the node becomes inactive.

\begin{figure}[htbp]
    \centering
        \includegraphics[width=0.9\linewidth]{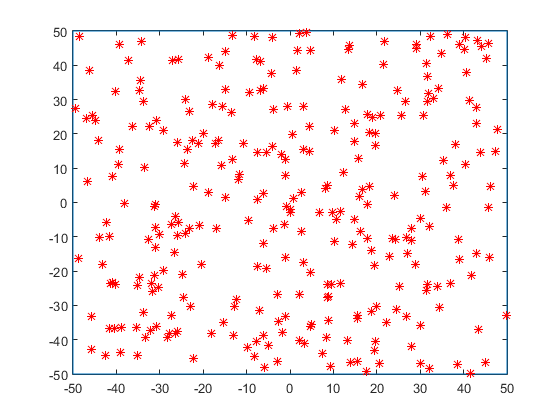}
     \caption{Network Environment at the initial state}
  \label{fig:fig1}
\end{figure}

This affects network coverage and density, which results in unbalanced energy consumption in different areas which inevitably results in electing low-energy nodes as CH. In contrast, in heterogeneous IoT networks it is possible to elect high energy nodes to play the CH role. Figure \ref{fig:fig1} illustrates the network and node distribution in the initial phase.

% \begin{figure}[htbp]
% \centerline{\includegraphics{figures/fig1.jpg}}
% \caption{Example of a figure caption.}
% \label{fig}
% \end{figure}

\section{Proposed Approach}
Clustering should be done so that highly energetic nodes that are denser in terms of position receive more packets from the neighboring nodes. Therefore, it is necessary to calculate the residual energy values of the nodes so that the re-clustering operation is performed according to the states and energy statuses of the nodes. The proposed statistically-based method is a re-modification of CH selection methods based on a change in the role of the CH through calculation of the amount of real energy in joules divided by the average of the surroundings and distribution of the energy consumption value among the initial highly-energetic nodes.In the first round, nodes with high real energy values rather than terms of percentage are selected. Then, as mean neighbor energy status is used, the energy consumption value is distributed so that high-energy nodes play a role in the sending of packets to the gateways. Therefore, energy consumption in the cluster is reduced. With these parameters taken into account, the proposed model for CH selection is shown as in Eq. (1).

\begin{multline}
    CH_{\text{prob}}=\\
    C_{\text{prob}}\times{\frac{E_{\text{rest current node in }j}}{M_{E_{\text{rest neighbor nodes in }j}}}}\times{\frac{E_{\text{rest}}}{\sqrt{\frac{1}{N-1}}\sum_{i=1}^{N}(E_i-M)^2}}
    \label{eq1}
\end{multline}

In Eq. \ref{eq1}, \(CH_{\text{prob}}\) is the probability of the \(CH\) role, which is based on the following factors. \(C_{\text{prob}}\) has a fixed probability value of 5\%. \(E_{\text{rest}}\) is the amount of node energy that is considered in joules. \(M\) is mean neighboring node energy, which appears in the denominator of the average residual energy function of the target node, also calculated in joules. In fact, a comparison is made in this part of our algorithm between the residual energy of the node and the residual energy state of the surrounding nodes in a cluster.is the amount of node energy that is considered in joule. In the next term of the equation, the residual energy of the desired node is again considered in joules, but sample deviation is used this time for calculation of the variance of the desired node. In the deviation section, the criterion \(N\) is the number of surrounding nodes, \(E_i\) is the residual energy of each node, which is subtracted from the mean for the neighboring nodes. 

To calculate the relative variance of residual energy in the network, we used sample deviation for each cluster; that is, a comparison is made between the residual energy of the node and the sample deviation of the average energy of the remaining neighbour nodes around. In fact, the variance of residual energy of the node is calculated relative to its neighboring nodes. A positive value for the standard deviation means that the examined node has more energy than the average energy of its neighbour nodes, and therefore, will get elected as the CH. In Eq (2), another term will be added, which will serve as a criterion for selecting the CH based on the conditions. On that basis, the number of repetitions in the competition round with other nodes will also be considered.

\begin{multline}
   CH_{\text{prob}}=\\C_{\text{prob}}\times{\frac{E_{\text{rest current node in }j}}{M_{E_{\text{rest neighbor nodes in }j}}}}\times{\frac{E_{\text{rest}}}{\sqrt{\frac{1}{N-1}}\sum_{i=1}^{N}(E_i-M)^2}}
    \label{eq2}\times\\{min(\frac{\text{node degree}}{D_{\text{avg}}},1)}
\end{multline}

In Eq. (2), the three parts of the previous formula are repeated, except that in this step, it is necessary for specification of the protocol of choice at the node and its selection as a CH to consider its position and the density of the surrounding nodes. In the third part of the formula, nodes with more sparse surroundings are selected as CHs, and their data need to be transferred. ${D_{\text{avg}}}$ is the average density of the surrounding nodes, calculated as in Eq. (3).

\begin{equation}
   {{D_{\text{avg}}=\ {\pi}R^2}}\times{\frac{Num_{\text{Devices }}}{Area}}
\end{equation}

In Eq. (3), $R$ is the communication radius of the node, ${Num_{\text{Devices }}}$ is the number of battery-powered devices in network, and $Area$ is the operation environment of sensors and devices. The value is $1$ when the node can be elected as the final CH. Where the node is subjected to calculation, there are three possible cases, as given in Eq. (4).

\begin{multline}
\begin{cases}
C_1:\text{Random}(0,1)\leq{Cprob} & [1]\\
C_2:\frac{E_{\text{rest in j}}}{M_{E_{\text{rest other nodes in neighbourhood in j}}}}\times\\{\frac{E_{\text{rest}}}{\sqrt{\frac{1}{N-1}}\sum_{i=1}^{N}(E_i-M)^2}}\geq{1} & [2] \\\
C_3:\frac{\text{Node Degree}}{D_{\text{avg}}}\geq{1} & [3]
\label{eq:conditions}
\end{cases}
\end{multline}

If two of the three conditions in Eq. (4) are met, the node will be selected as the final node. If the
node in question is unable to communicate with any of the nodes, it prepares itself for the transfer as
the final CH without having to participate in the competition. Table II summarizes represents a truth table of possible scenarios\cite{s17071574}.

\begin{table}[htbp]
\label{tb:CH_statuses1}
\caption{Cluster head statuses}
\begin{center}
\begin{tabular}{M{1.5cm}|M{1.5cm}|M{1.5cm}|M{1.5cm}}
\hline
\rowcolor{gray!20}
    \textbf{\(C_1\)} & \textbf{\(C_2\)} & \textbf{\(C_3\)} & \textbf{\(C_4\)} \\
    \hline
    \hline
    1 & 1 & 1 & Final \\
    0 & 1 & 1 & Final \\
    1 & 0 & 1 & Final \\
    1 & 1 & 0 & Final \\
    1 & 0 & 0 & Tentative \\
\end{tabular}
\end{center}
\end{table}

The simulation parameters are indicated in Table III. Network communication interfaces are NAI1...NAI5 which includes Wifi, Bluetooth, ZigBee, LTE and NB-LTE and each one has its own characteristics. The detailed information for these communication technologies and power consumption models are mentioned in \cite{s17071574}. 

\begin{table}[htbp]
\caption{Simulation Parameters}
\begin{center}
\begin{tabular}{c|c|c p{5cm}}
\hline 
\rowcolor{gray!20}
\textbf{Type} & \textbf{Parameters} & \textbf{Values}\\
\hline \hline
   Area  & Network Area & From (0,0) - (100, 100) \\
   Number of nodes & N & 300  \\w7
   Initial energy & Energy & Heterogeneous  \\
   Transmission range  & R & various  \\
   Gateway & Sink & At (50, 175)  \\
   Default cluster radious &  & 25m  \\
   Data packet size &  & 100 bytes  \\
   Broadcast packet size &  & 25 bytes  \\
   Data header size &  & 25 bytes  \\
   Each round &  & 5 TDMA frames  \\
   CH data compress rate &  & 0.8  \\
   
   Duration &  & 1000 rounds  \\
   Interfaces &  & NAI1-NAI5  \\
   
\end{tabular}
\end{center}
\end{table}

\section{Performance Evaluation}
In recent years, several works aim to simulate various layers of cloud computing stack: IoT \cite{10.5555/1941192}, \cite{4116633} edge \cite{8654084}, \cite{ Yousefpour_2019}, fog \cite{https://doi.org/10.1002/spe.2509}, and cloud-based scenarios \cite{khan2021perfsim} by considering different levels of system details and complexities. However, in the context of clustering algorithms, the majority of papers used MATLAB as their base simulation platform \cite{matlab} due to the statistical analysis nature of their evaluations. For the same reason, to evaluate our method we also used MATLAB as the base simulation platform and all conditions, including network size, simulation area, and number of nodes, as well as network distribution are assumed to be the same in all compared algorithms. The average of ten runs in the simulation process was calculated to obtain the final results.

In this study, we named the result of the proposed algorithm HetEng. It was compared statistically with different algorithms, including HEED, LEACH, BEE, BEEM, and Smart-BEEM. As initial energy levels were generated randomly in the simulation environment, the running phase was assumed to be the same in all the algorithms. For large areas, the proposed method exhibited an increase in network lifetime as compared to the other algorithms. It was also observed that network lifespan was generally longer in networks with more nodes than in ones with fewer nodes. The results indicated an increase in the number of alive nodes and residual energy in the proposed algorithm as compared to the others. In the large networks, however, network coverage was much closer to network lifetime, because the energy consumption in the network was significantly higher.

\begin{figure}[htbp]
    \centering
        \includegraphics[width=1\linewidth]{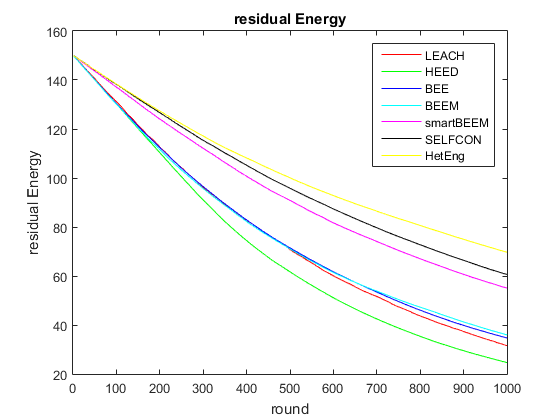}
     \caption{Comparison of residual energy among different algorithms}
  \label{fig:fig2}
\end{figure}

In Fig. \ref{fig:fig2}, the superiority of the proposed algorithm over all the compared algorithms can be observed. On that basis, an average 3\% improvement over the Smart-BEEM algorithm has occurred in the residual energy of the nodes, and the slope is closer to linear than those of the other algorithms, which has resulted from the closer-to-normal distribution of energy consumption. Moreover, the HEED and LEACH algorithms exhibited steeper slopes, in that order, which indicated lower amounts of energy remaining after 1000 rounds.

\begin{figure}[hbtp]
    \centering
        \includegraphics[width=1\linewidth]{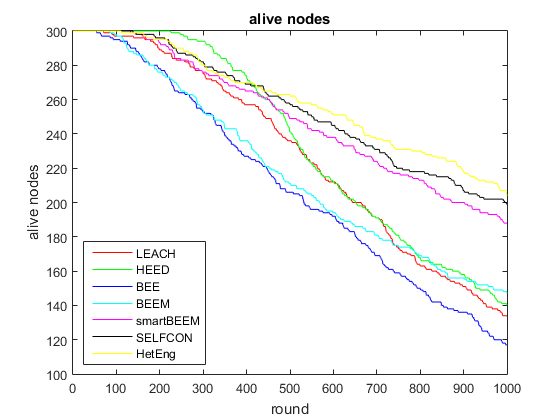}
     \caption{Alive nodes state in the network}
  \label{fig:fig3}
\end{figure}

Fig. \ref{fig:fig3}, shows how the normal distribution of the CH role and the use of high-energy nodes to send packets resulted in the distribution of energy consumption over the network. As depicted in figure  \ref{fig:fig3}, the number of alive nodes exhibited an improvement by 6.6\% on average in the proposed algorithm with respect to that in Smart-BEEM. The slope of the diagram in HetEng is closer to normal and linear than in the compared algorithms, which demonstrates the slower energy discharge among network nodes.

\begin{figure}[htbp]
    \centering
        \includegraphics[width=1\linewidth]{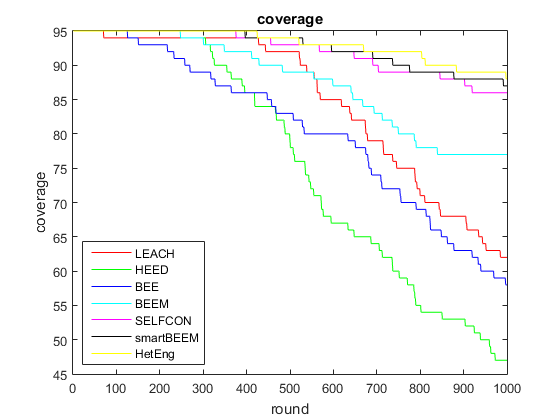}
     \caption{Network coverage among different algorithms}
  \label{fig:fig4}
\end{figure}

As indicated in Fig. \ref{fig:fig4}, in term of network coverage, the proposed algorithm exhibited a slight improvement in some scenarios as it depends on different factors e.g. nodes distribution and their initial energy state in the scenario, however, its performance was equal to that of Smart-BEEM and SelfCon on average, resulting from the use of five communication protocols in all the compared algorithms, which greatly affects network coverage. The diagram exhibited more normal slopes for Smart-BEEM, SelfCon, and HetEng than other compared algorithms.

\begin{figure}[htbp]
    \centering
        \includegraphics[width=1\linewidth]{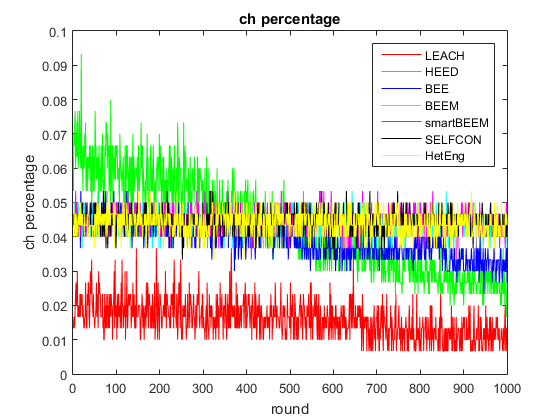}
     \caption{CH percentage in each algorithm}
  \label{fig:fig5}
\end{figure}
Fig. \ref{fig:fig5}, demonstrates the probability of selecting a CH among the nodes. As can be seen, the HEED algorithm exhibited the largest number of jumps and changes at the beginning of the competition, and the probability constantly decreased over time with the loss of alive nodes. The five algorithms BEE, BEEM, Smart-BEEM, SelfCon, and HetEng were closer to linear, and exhibited almost equal values of probability of the role.

\begin{figure}[htbp]
    \centering
        \includegraphics[width=1\linewidth]{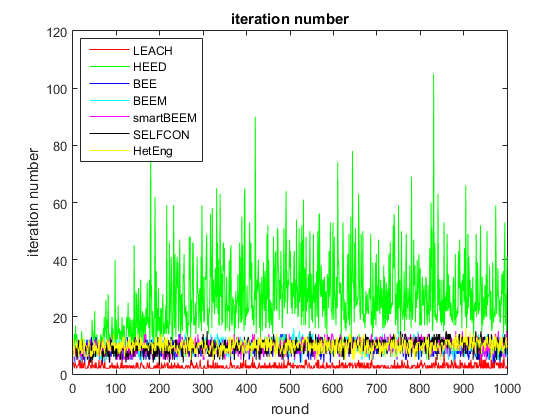}
     \caption{Iteration Status}
  \label{fig:fig6}
\end{figure}
Fig. \ref{fig:fig6}, shows the importance of reducing the number of iterations in the selection of the CH in the network and its impact on reduction of network overhead. As indicated, the HEED algorithm exhibited the largest number of rounds and of fluctuations due to the greedy mechanism of selecting the cluster head and the focus on distribution of the role among all the nodes (including those with low energy). HetEng reduced the iteration number by 1\% on average in comparison with Smart-BEEM and SelfCon on average. It should be noted that the Leach algorithm exhibited the smallest number of competition rounds and the optimal mode among all the algorithms due to the application of random selection instead of competition rounds.

\section{Conclusion and Future Works}

In this paper, we proposed a distributed energy-efficient cluster head (CH) selection scheme in low-power clustered IoT networks to support heterogeneity and detect the most optimum nodes with high energy to play the CH role. Despite the many advancements that have been made in the field of hardware and downsizing technologies, energy consumption, network longevity, and, most importantly, network coverage are considered as the main challenges in such networks. Many of the challenges that exist in all low-power networks, such as wireless sensor networks, are still there in IoT networks, where they are even severer due to the much more complex scenarios. In the proposed algorithm, real energy variance was calculated using sample deviation, and the nodes were compared to those surrounding them in the cluster, indicating a high level of intelligence, which makes it possible to manage the network in terms of longevity, energy consumption, and network coverage under different conditions. We reduced the number of iterations by 1\%, and used average values to distribute energy consumption among high-energy nodes, which resulted in: 1- reduction of energy consumption by 6.6\% over the whole network rather than at each individual node and 2- improvement of coverage with respect to that in the compared algorithms through prevention of rapid depletion of energy at nodes with low energy. Finally, there are issues that need to be addressed for improvement of the present study in more realistic environments: (1) mobility, which is an important factor in IoT networks, and (2) Quality of Service (QoS).

\section*{Acknowledgment}

The authors would like to express their highest gratitude to Mr. Michel Gokan Khan from Karlstad University in Sweden for his genuine support and constructive feedback.

% \section*{References}

% Please number citations consecutively within brackets \cite{6785450}. The 
% sentence punctuation follows the bracket \cite{b2}. Refer simply to the reference 
% number, as in \cite{b3}---do not use ``Ref. \cite{b3}'' or ``reference \cite{b3}'' except at 
% the beginning of a sentence: ``Reference \cite{b3} was the first $\ldots$''

% Number footnotes separately in superscripts. Place the actual footnote at 
% the bottom of the column in which it was cited. Do not put footnotes in the 
% abstract or reference list. Use letters for table footnotes.

% Unless there are six authors or more give all authors' names; do not use 
% ``et al.''. Papers that have not been published, even if they have been 
% submitted for publication, should be cited as ``unpublished'' \cite{b4}. Papers 
% that have been accepted for publication should be cited as ``in press'' \cite{b5}. 
% Capitalize only the first word in a paper title, except for proper nouns and 
% element symbols.

% For papers published in translation journals, please give the English 
% citation first, followed by the original foreign-language citation \cite{b6}.

\bibliographystyle{IEEEtran}
\bibliography{IEEEabrv,bib}

\begin{thebibliography}{10}
\providecommand{\url}[1]{#1}
\csname url@samestyle\endcsname
\providecommand{\newblock}{\relax}
\providecommand{\bibinfo}[2]{#2}
\providecommand{\BIBentrySTDinterwordspacing}{\spaceskip=0pt\relax}
\providecommand{\BIBentryALTinterwordstretchfactor}{4}
\providecommand{\BIBentryALTinterwordspacing}{\spaceskip=\fontdimen2\font plus
\BIBentryALTinterwordstretchfactor\fontdimen3\font minus
  \fontdimen4\font\relax}
\providecommand{\BIBforeignlanguage}[2]{{%
\expandafter\ifx\csname l@#1\endcsname\relax
\typeout{** WARNING: IEEEtran.bst: No hyphenation pattern has been}%
\typeout{** loaded for the language `#1'. Using the pattern for}%
\typeout{** the default language instead.}%
\else
\language=\csname l@#1\endcsname
\fi
#2}}
\providecommand{\BIBdecl}{\relax}
\BIBdecl

\bibitem{amiri-azimi-acmhealth20}
\BIBentryALTinterwordspacing
D.~Amiri, A.~Anzanpour, I.~Azimi, M.~Levorato, P.~Liljeberg, N.~Dutt, and A.~M.
  Rahmani, ``Context-aware sensing via dynamic programming for edge-assisted
  wearable systems,'' \emph{ACM Trans. Comput. Healthcare}, vol.~1, no.~2, Mar.
  2020. [Online]. Available: \url{https://doi.org/10.1145/3351286}
\BIBentrySTDinterwordspacing

\bibitem{7976279}
L.~{Xu}, R.~{Collier}, and G.~M.~P. {O’Hare}, ``A survey of clustering
  techniques in wsns and consideration of the challenges of applying such to 5g
  iot scenarios,'' \emph{IEEE Internet of Things Journal}, vol.~4, no.~5, pp.
  1229--1249, 2017.

\bibitem{9246553}
A.~{Shahraki}, A.~{Taherkordi}, {Haugen}, and F.~{Eliassen}, ``A survey and
  future directions on clustering: From wsns to iot and modern networking
  paradigms,'' \emph{IEEE Transactions on Network and Service Management}, pp.
  1--1, 2020.

\bibitem{s17071574}
\BIBentryALTinterwordspacing
L.~Xu, G.~M.~P. O’Hare, and R.~Collier, ``A smart and balanced
  energy-efficient multihop clustering algorithm (smart-beem) for mimo iot
  systems in future networks †,'' \emph{Sensors}, vol.~17, no.~7, 2017.
  [Online]. Available: \url{https://www.mdpi.com/1424-8220/17/7/1574}
\BIBentrySTDinterwordspacing

\bibitem{s21030873}
\BIBentryALTinterwordspacing
M.~S. Batta, H.~Mabed, Z.~Aliouat, and S.~Harous, ``A distributed multi-hop
  intra-clustering approach based on neighbors two-hop connectivity for iot
  networks,'' \emph{Sensors}, vol.~21, no.~3, 2021. [Online]. Available:
  \url{https://www.mdpi.com/1424-8220/21/3/873}
\BIBentrySTDinterwordspacing

\bibitem{5679898}
P.~{Saini} and A.~K. {Sharma}, ``E-deec- enhanced distributed energy efficient
  clustering scheme for heterogeneous wsn,'' in \emph{2010 First International
  Conference On Parallel, Distributed and Grid Computing (PDGC 2010)}, 2010,
  pp. 205--210.

\bibitem{LI2017124}
\BIBentryALTinterwordspacing
Q.~Li, S.~P. Gochhayat, M.~Conti, and F.~Liu, ``Energiot: A solution to improve
  network lifetime of iot devices,'' \emph{Pervasive and Mobile Computing},
  vol.~42, pp. 124--133, 2017. [Online]. Available:
  \url{https://www.sciencedirect.com/science/article/pii/S1574119217301219}
\BIBentrySTDinterwordspacing

\bibitem{MALAMBO20191}
\BIBentryALTinterwordspacing
L.~Malambo, S.~Popescu, D.~Horne, N.~Pugh, and W.~Rooney, ``Automated detection
  and measurement of individual sorghum panicles using density-based clustering
  of terrestrial lidar data,'' \emph{ISPRS Journal of Photogrammetry and Remote
  Sensing}, vol. 149, pp. 1--13, 2019. [Online]. Available:
  \url{https://www.sciencedirect.com/science/article/pii/S0924271618303502}
\BIBentrySTDinterwordspacing

\bibitem{10.1007/s11276-018-1679-2}
\BIBentryALTinterwordspacing
D.~R. Edla, M.~C. Kongara, and R.~Cheruku, ``Sce-pso based clustering approach
  for load balancing of gateways in wireless sensor networks,'' \emph{Wirel.
  Netw.}, vol.~25, no.~3, p. 1067–1081, Apr. 2019. [Online]. Available:
  \url{https://doi.org/10.1007/s11276-018-1679-2}
\BIBentrySTDinterwordspacing

\bibitem{https://doi.org/10.1002/spe.2797}
\BIBentryALTinterwordspacing
C.~Iwendi, P.~K.~R. Maddikunta, T.~R. Gadekallu, K.~Lakshmanna, A.~K. Bashir,
  and M.~J. Piran, ``A metaheuristic optimization approach for energy
  efficiency in the iot networks,'' \emph{Software: Practice and Experience},
  vol. n/a, no. n/a. [Online]. Available:
  \url{https://onlinelibrary.wiley.com/doi/abs/10.1002/spe.2797}
\BIBentrySTDinterwordspacing

\bibitem{GAJJAR2016235}
\BIBentryALTinterwordspacing
S.~Gajjar, M.~Sarkar, and K.~Dasgupta, ``Famacrow: Fuzzy and ant colony
  optimization based combined mac, routing, and unequal clustering cross-layer
  protocol for wireless sensor networks,'' \emph{Applied Soft Computing},
  vol.~43, pp. 235--247, 2016. [Online]. Available:
  \url{https://www.sciencedirect.com/science/article/pii/S1568494616300709}
\BIBentrySTDinterwordspacing

\bibitem{7502988}
M.~Ren, L.~Khoukhi, H.~Labiod, J.~Zhang, and V.~Veque, ``A new mobility-based
  clustering algorithm for vehicular ad hoc networks (vanets),'' in \emph{NOMS
  2016 - 2016 IEEE/IFIP Network Operations and Management Symposium}, 2016, pp.
  1203--1208.

\bibitem{926982}
W.~R. {Heinzelman}, A.~{Chandrakasan}, and H.~{Balakrishnan},
  ``Energy-efficient communication protocol for wireless microsensor
  networks,'' in \emph{Proceedings of the 33rd Annual Hawaii International
  Conference on System Sciences}, 2000, pp. 10 pp. vol.2--.

\bibitem{https://doi.org/10.1049/iet-wss.2017.0099}
\BIBentryALTinterwordspacing
T.~M. Behera, U.~C. Samal, and S.~K. Mohapatra, ``Energy-efficient modified
  leach protocol for iot application,'' \emph{IET Wireless Sensor Systems},
  vol.~8, no.~5, pp. 223--228, 2018. [Online]. Available:
  \url{https://ietresearch.onlinelibrary.wiley.com/doi/abs/10.1049/iet-wss.2017.0099}
\BIBentrySTDinterwordspacing

\bibitem{1347100}
O.~Younis and S.~Fahmy, ``Heed: a hybrid, energy-efficient, distributed
  clustering approach for ad hoc sensor networks,'' \emph{IEEE Transactions on
  Mobile Computing}, vol.~3, no.~4, pp. 366--379, 2004.

\bibitem{6878886}
L.~{Xu}, G.~M.~P. {O'Hare}, and R.~{Collier}, ``A balanced energy-efficient
  multihop clustering scheme for wireless sensor networks,'' in \emph{2014 7th
  IFIP Wireless and Mobile Networking Conference (WMNC)}, 2014, pp. 1--8.

\bibitem{8645769}
A.~Mandeh, K.~Khamforoosh, and V.~Maihami, ``Data fusion in wireless sensor
  networks using fuzzy systems,'' \emph{International Journal of Computer
  Applications}, vol. 125, pp. 31--36, 09 2015.

\bibitem{khamforoosh2011clustered}
K.~Khamforoosh, ``Clustered balanced minimum spanning tree for routing and
  energy reduction in wireless sensor networks,'' in \emph{2011 IEEE Symposium
  on Wireless Technology and Applications (ISWTA)}.\hskip 1em plus 0.5em minus
  0.4em\relax IEEE, 2011, pp. 56--59.

\bibitem{SADEK2018166}
\BIBentryALTinterwordspacing
R.~A. Sadek, ``Hybrid energy aware clustered protocol for iot heterogeneous
  network,'' \emph{Future Computing and Informatics Journal}, vol.~3, no.~2,
  pp. 166--177, 2018. [Online]. Available:
  \url{https://www.sciencedirect.com/science/article/pii/S2314728818300163}
\BIBentrySTDinterwordspacing

\bibitem{10.5555/1941192}
K.~Wehrle, M.~Gnes, and J.~Gross, \emph{Modeling and Tools for Network
  Simulation}, 1st~ed.\hskip 1em plus 0.5em minus 0.4em\relax Springer
  Publishing Company, Incorporated, 2010.

\bibitem{4116633}
F.~Osterlind, A.~Dunkels, J.~Eriksson, N.~Finne, and T.~Voigt, ``Cross-level
  sensor network simulation with cooja,'' in \emph{Proceedings. 2006 31st IEEE
  Conference on Local Computer Networks}, 2006, pp. 641--648.

\bibitem{8654084}
R.~Buyya and S.~N. Srirama, \emph{Modeling and Simulation of Fog and Edge
  Computing Environments Using iFogSim Toolkit}, 2019, pp. 433--465.

\bibitem{Yousefpour_2019}
\BIBentryALTinterwordspacing
A.~Yousefpour, C.~Fung, T.~Nguyen, K.~Kadiyala, F.~Jalali, A.~Niakanlahiji,
  J.~Kong, and J.~P. Jue, ``All one needs to know about fog computing and
  related edge computing paradigms: A complete survey,'' \emph{Journal of
  Systems Architecture}, vol.~98, p. 289–330, Sep 2019. [Online]. Available:
  \url{http://dx.doi.org/10.1016/j.sysarc.2019.02.009}
\BIBentrySTDinterwordspacing

\bibitem{https://doi.org/10.1002/spe.2509}
\BIBentryALTinterwordspacing
H.~Gupta, A.~Vahid~Dastjerdi, S.~K. Ghosh, and R.~Buyya, ``ifogsim: A toolkit
  for modeling and simulation of resource management techniques in the internet
  of things, edge and fog computing environments,'' \emph{Software: Practice
  and Experience}, vol.~47, no.~9, pp. 1275--1296, 2017. [Online]. Available:
  \url{https://onlinelibrary.wiley.com/doi/abs/10.1002/spe.2509}
\BIBentrySTDinterwordspacing

\bibitem{khan2021perfsim}
\BIBentryALTinterwordspacing
M.~Gokan~Khan, J.~Taheri, A.~Al-Dulaimy, and A.~Kassler, ``Perfsim: {A}
  performance simulator for cloud native computing,'' \emph{CoRR}, vol.
  abs/2103.08983, 2021. [Online]. Available:
  \url{https://arxiv.org/abs/2103.08983}
\BIBentrySTDinterwordspacing

\bibitem{matlab}
\BIBentryALTinterwordspacing
Matlab. [Online]. Available: \url{https://www.mathworks.com/}
\BIBentrySTDinterwordspacing

\end{thebibliography}

\end{document}